\documentclass[12pt]{article}

\usepackage{amssymb, amsmath, amsthm, amsfonts, graphicx, color, bm, float} 
\usepackage[margin=1in]{geometry}
\usepackage{hyperref}
\usepackage{chicago}
\usepackage{setspace} 
\usepackage{graphicx, subfig, epstopdf, enumitem}
\usepackage{caption}

\usepackage{algorithm, algpseudocode}

\newfloat{algorithm}{htbp}{loa}
\floatname{algorithm}{Algorithm}

\title{ABC Samplers}

\author{Y. Fan\footnote{School of Mathematics and Statistics, University of New South Wales, Australia.}\:\:  and S. A. Sisson$^{*}$}

\begin{document}

\maketitle

\section{Introduction}

{\em Approximate Bayesian computation} (ABC) is a phrase that describes a collection of methods and algorithms designed to perform a Bayesian analysis using an approximation to the true posterior distribution, when the likelihood function implied by the data generating process is computationally intractable.
For observed data $y_{obs}\in\mathcal{Y}$, the likelihood function $p(y|\theta)$ depends on a vector of model parameters $\theta\in \Theta$, from which prior beliefs $\pi(\theta)$ may be updated into posterior beliefs $\pi(\theta|y_{obs})\propto p(y_{obs}|\theta)\pi(\theta)$ via Bayes' theorem.
In the standard ABC framework (see e.g. \shortciteNP{sisson+fb17}, this volume) the ABC approximation to $\pi(\theta|y_{obs})$ is given by
\begin{eqnarray}
\label{eqn:abc-post}
	\pi_{ABC}(\theta|s_{obs})\propto\int K_h(\|s-s_{obs}\|)p(s|\theta)\pi(\theta)ds,
\end{eqnarray}
where $K_h(u)=K(u/h)/h$ is a standard kernel density function with scale parameter $h>0$, $\|\cdot\|$ is an appropriate distance metric (e.g. Euclidean or Mahalanobis distance), $p(s|\theta)$
is the (intractable) likelihood function of the low-dimensional vector of summary statistics $s=S(y)$ implied by $p(y|\theta)$, and $s_{obs}=S(y_{obs})$. 
Defining $K_h(\|s-s_{obs}\|)\rightarrow \delta_{s_{obs}}(s)$ as $h\rightarrow 0$, 
where $\delta_Z(z)$ denotes the Dirac measure, defined as $\delta_Z(z)=1$ if $z \in Z$ 
and $\delta_Z(z)=0$ otherwise,
then as a result 
\[
	\lim_{h\rightarrow 0}\pi_{ABC}(\theta|s_{obs})\propto \int \delta_{s_{obs}}(s)p(s|\theta)\pi(\theta)ds = p(s_{obs}|\theta)\pi(\theta)\propto \pi(\theta|s_{obs}).
\]
Accordingly, $\pi_{ABC}(\theta|s_{obs})$ provides an approximation to the partial posterior $\pi(\theta|s_{obs})$, which becomes more accurate as $h$ gets small.
If the summary statistics $s$ are sufficient for $\theta$, then $\pi(\theta|s_{obs})$ will equal $\pi(\theta|y_{obs})$, and so, for small $h$,
the ABC posterior approximation $\pi_{ABC}(\theta|s_{obs})$ will be a good approximation of the true posterior. If either $s$ is not sufficient, or $h$ is not small, then the ABC posterior approximation $\pi_{ABC}(\theta|s_{obs})$ will be of the form (\ref{eqn:abc-post}). 

In terms of drawing samples from the approximate posterior $\pi_{ABC}(\theta|s_{obs})$, the choice of summary statistics $s=S(y)$ is typically  considered known, and interest is then in sampling from $\pi_{ABC}(\theta|s_{obs})$, for a specific and low value of $h$, as efficiently as possible. The more efficient the simulation procedure, the further $h$ can be lowered within the sampling framework, resulting in samples from a more accurate approximation of $\pi(\theta|s_{obs})$.

In this chapter we survey the various forms of ABC algorithms that have been developed to sample from $\pi_{ABC}(\theta|s_{obs})$. These have broadly followed the familiar Monte Carlo classes of algorithms, including rejection and importance sampling, Markov chain Monte Carlo (MCMC), and sequential Monte Carlo (SMC) based algorithms. While each of these classes have  their ABC-specific implementations and  characteristics, in general they target the joint distribution of parameter vector $\theta$ and summary statistic $s$ given by
\begin{eqnarray}
\label{eqn:abc-joint}
	\pi_{ABC}(\theta,s|s_{obs})\propto K_h(\|s-s_{obs}\|)p(s|\theta)\pi(\theta).
\end{eqnarray}
By noting that (\ref{eqn:abc-post}) is obtained from (\ref{eqn:abc-joint}) by integrating over $s$ (i.e. $\pi_{ABC}(\theta|s_{obs})=\int \pi_{ABC}(\theta,s|s_{obs})ds$), samples from $\pi_{ABC}(\theta|s_{obs})$ can  be obtained by first drawing samples from (\ref{eqn:abc-joint}) and then discarding the marginal $s$ values.

An alternative, but  related Monte Carlo approach is based on sampling from (\ref{eqn:abc-post}) directly, by 
obtaining an unbiased and non-negative estimate of the ABC posterior distribution function, such as
\[
	\hat{\pi}_{ABC}(\theta|s_{obs}) \propto \frac{\pi(\theta)}{T}\sum_{t=1}^{T} K_h(\|s{(t)}-s_{obs}\|),
\]
where $s{(1)},\ldots,s{(T)}\sim p(s|\theta)$ are samples from the intractable model given $\theta$, and then using this estimate in place of $\pi_{ABC}(\theta|s_{obs})$ within a standard Monte Carlo algorithm (e.g. \shortciteNP{delmoral+dj12}). This approach falls within the family of {\em pseudo-marginal} Monte Carlo methods \cite{beaumont03,andrieu+r09}, a more general class of likelihood-free samplers that has gained popularity outside of the ABC setting. For a detailed and in-depth discussion of the connections between ABC-MCMC algorithms and pseudo-marginal MCMC methods see \shortciteN{andrieu+sv17} (this volume).

\section{Rejection and importance sampling}

\subsection{Rejection sampling}
\label{sec:rej}

The earliest ABC samplers (e.g. \shortciteNP{tavare+bgd97,pritchard+spf99}) were basic rejection sampling algorithms.
Under the standard rejection sampling framework (e.g.  \citeNP{rip87,liu01}) interest is in 
obtaining samples from some target distribution $f(\theta)=Z^{-1}\tilde{f}(\theta)$, which is known up to a normalising constant $Z=\int_{-\infty}^\infty \tilde{f}(\theta)d\theta$. The standard rejection sampling algorithm obtains draws $\theta'\sim g(\theta)$ from a sampling density $g(\theta)$ from which it is trivial to sample,   such that 
$$\tilde{f}(\theta)\leq  M g(\theta),$$
for all $\theta$ and some positive constant $M>0$.
The draws $\theta'$ are then accepted as  independent samples from the target density $f(\theta)$ with probability $\frac{\tilde{f}(\theta')}{Mg(\theta')}$. 
To see that the above procedure is correct, for simplicity consider the case in which $\theta$ is univariate, with the extension to multivariate $\theta$ being straightforward.  Define $A$ as the event that a sample $\theta'$ from $g(\theta)$ is accepted. Then, the overall acceptance rate of the algorithm is
$$\mbox{Pr}(A) = \int \frac{\tilde{f}(\theta)}{Mg(\theta)}g(\theta)d\theta= \frac{1}{M}\int \tilde{f}(\theta) d\theta = \frac{Z}{M},$$
and hence the distribution of accepted draws is
$$\mbox{Pr}(\theta'\leq\theta | A) = \frac{\mbox{Pr}(\theta'\leq\theta, A)}{\mbox{Pr}(A)}=\frac{\int_{-\infty}^\theta \frac{\tilde{f}(\theta')}{Mg(\theta')}g(\theta')d\theta'}{P(A)}= F(\theta),$$
as required, where $F(\theta)=\int_{-\infty}^\theta f(z)dz$ is the distribution function associated with $f(\theta)$. 
The efficiency of the algorithm is associated with the value of $M$, with smaller  values of $M$ (subject to $\tilde{f}(\theta)\leq  M g(\theta)$, $\forall \theta$) corresponding to more efficient samplers. That is, for fixed $g(\theta)$ the optimum choice is $M=\max_\theta \frac{\tilde{f}(\theta)}{g(\theta)}$. Good choice of the sampling distribution $g(\theta)$, e.g. to approximate $f(\theta)$, can result in smaller values of $M$.

The ABC version of the rejection sampler was discussed in \shortciteN{sisson+fb17} (this volume), which we reproduce here as Algorithm 1.

\begin{table}[tbh]
\caption{\bf Algorithm 1: ABC Rejection Sampling}
\noindent {\it Inputs:}
\begin{itemize}
\item A target posterior density $\pi(\theta|y_{obs})\propto p(y_{obs}|\theta)\pi(\theta)$, consisting of a prior distribution $\pi(\theta)$ and a procedure for generating data under the model  $p(y_{obs}|\theta)$.
\item A proposal density $g(\theta)$, with $g(\theta)>0$ if $\pi(\theta|y_{obs})>0$.
\item An integer $N>0$.
\item A kernel function $K_h(u)$ and scale parameter $h>0$.
\item A low dimensional vector of summary statistics $s=S(y)$.
\\
\end{itemize}

\noindent {\it Sampling:}\\
\noindent For $i=1, \ldots, N$:
\begin{enumerate}
\item \label{chapter3:alg:ABC-rejectionSS:step1} Generate $\theta^{(i)}\sim g(\theta)$ from sampling density $g$.
\item Generate $y^{(i)}\sim p(y|\theta^{(i)})$ from the model.
\item Compute summary statistic $s^{(i)}=S(y^{(i)})$.
\item Accept $\theta^{(i)}$ with probability $\frac{K_h(\|s^{(i)}-s_{obs}\|)\pi(\theta^{(i)})}{Mg(\theta^{(i)})}$
	where $M\geq K_h(0)\max_\theta\frac{\pi(\theta)}{g(\theta)}$.\\ Else go to \ref{chapter3:alg:ABC-rejectionSS:step1}.
\\
\end{enumerate}

\noindent {\it Output:}\\
A set of parameter vectors $\theta^{(1)},\ldots,\theta^{(N)}$ $\sim$ $\pi_{ABC}(\theta|s_{obs})$.
\end{table}

Originally developed by \shortciteN{pritchard+spf99} following earlier ideas by \shortciteN{tavare+bgd97}, the ABC rejection sampling algorithm is typically described heuristically as follows: for the candidate parameter vector $\theta'\sim g(\theta)$, a dataset $y'$ is generated from the (intractable) generative model and summary statistics $s'=S(y')$ computed.
If the simulated and observed datasets are similar (in some manner), so that $s'\approx s_{obs}$, then $\theta'$ could credibly have generated the observed data under the given model, and so $\theta'$ is retained and forms part of the sample from the ABC posterior distribution $\pi(\theta |s_{obs})$. 
Conversely, if $s'$ and $s_{obs}$ are dissimilar, then $\theta'$ is unlikely to have generated the observed data for this model, and so $\theta'$ is discarded. 
The parameter vectors accepted under this approach offer support for $s_{obs}$ under the model, and so may be considered to be drawn approximately from the posterior distribution $\pi(\theta|s_{obs})$. In this manner, the evaluation of the likelihood $p(y_{obs} | \theta')$, essential to most Bayesian posterior simulation methods,  is replaced by an evaluation of the proximity of summaries of a simulated dataset $s'$ to the observed summaries $s_{obs}$.

More precisely, this algorithm targets $\pi_{ABC}(\theta,s|s_{obs})$ given by (\ref{eqn:abc-joint}), the joint distribution of parameter vector and summary statistic given $s_{obs}$. Accordingly the sampling distribution is also defined on this space as $g(\theta,s) = p(s|\theta)g(\theta)$, and the acceptance probability of the vector $(\theta,s)$ is then given by
\[
	\frac{\pi_{ABC}(\theta,s|s_{obs})}{M g(\theta,s)}
	\propto
	\frac{K_h(\|s-s_{obs}\|)p(s|\theta)\pi(\theta)}{Mp(s|\theta)g(\theta)}
	=
	\frac{K_h(\|s-s_{obs}\|)\pi(\theta)}{Mg(\theta)}.
\]
The normalising constant $M$ is similarly given by
\[
	M\geq\max_{s,  \theta} \frac{K_h( \|s- s_{obs}\|)p(s|\theta)\pi(\theta)}{p(s|\theta)g(\theta)}=K_h(0)\max_\theta\frac{\pi(\theta)}{g(\theta)}
\]
with $\max_sK_h(\|s-s_{obs}\|)=K_h(0)$ resulting from  the zero-mean, symmetry and (typically) unimodal characteristics of standard kernel density functions. Accordingly, the construction of the target and sampling distributions on the joint space $(\theta,s)$ results in the form of the acceptance probability and normalisation constant $M$ being free of intractable likelihood terms.
An example implementation of this algorithm is given in \shortciteN{sisson+fb17} (this volume).

\subsection{Importance sampling}
\label{sec:imp}

One down side of rejection sampling is the need to determine a near optimal value for the normalising constant $M$ in order to produce an efficient algorithm. Importance sampling is a procedure that, rather than calculating acceptance probabilities, avoids this by alternatively assigning the draw $\theta'\sim g(\theta)$ an (importance) weight $w(\theta')= f(\theta')/g(\theta')$. The weighted vector $\theta'$ is then a draw from $f(\theta)$,  and desired expectations under the target distribution $f$ are computed as weighted expectations under the importance sampling density $g$.

To see this, suppose that we are interested in estimating the expectation
\[
	E_f[h(\theta)] = \int h(\theta)f(\theta)d\theta. 
\] 
By defining $w(\theta)=f(\theta)/g(\theta)$ we have
\[
	E_g[w(\theta)h(\theta)] = \int w(\theta)h(\theta)g(\theta)d\theta 
	= \int h(\theta)f(\theta)d\theta = E_f[h(\theta)].
\]
In this manner we can then estimate the expectation as
\[
	E_f[h(\theta)] \approx \frac{1}{N}\sum_{i=1}^Nw^{(i)}h(\theta^{(i)}),
\]
where $w^{(i)}=w(\theta^{(i)})$, and where $\theta^{(i)}\sim g(\theta)$ are draws from $g$. In the more typical case where the target distribution is unnormalised, so that $f(\theta)=Z^{-1}\tilde{f}(\theta)$,  we can work with $\tilde{f}(\theta)$ only by defining $\tilde{w}(\theta)=\tilde{f}(\theta)/g(\theta)$ and then noting that
\begin{equation}
	\label{eqn:Zest}
	E_g[\tilde{w}(\theta)] = \int\tilde{w}(\theta)g(\theta)d\theta = \int \tilde{f}(\theta)d\theta = Z \approx\frac{1}{N}\sum_{i=1}^N\tilde{w}^{(i)},
\end{equation}
for $\theta^{(i)}\sim g(\theta)$, where $\tilde{w}^{(i)}=\tilde{w}(\theta^{(i)})$. As a result, the expectation $E_f[h(\theta)]$ may be approximated as
\begin{eqnarray}
	E_f[h(\theta)] & = & \int h(\theta)f(\theta) = \frac{1}{Z}\int h(\theta)\tilde{f}(\theta)\nonumber\\
	& = & \frac{1}{Z}\int\tilde{w}(\theta)h(\theta)g(\theta)d\theta = 
	\frac{E_g[\tilde{w}(\theta)h(\theta)]}{E_g[\tilde{w}(\theta)]}\nonumber\\
	&\approx& \frac{ \frac{1}{N}\sum_{i=1}^N\tilde{w}^{(i)}h(\theta^{(i)})}{\frac{1}{N}\sum_{i=1}^N\tilde{w}^{(i)}}
	= \sum_{i=1}^N W^{(i)}h(\theta^{(i)}),\label{eqn:expectation}
\end{eqnarray}
for $\theta^{(i)}\sim g(\theta)$, where $W^{(i)} = \tilde{w}^{(i)}/\sum_{j=1}^N \tilde{w}^{(j)}$ denotes normalised weights. This approximation is not unbiased due to the biased estimator of $1/Z$, although the bias becomes small as $N$ becomes large.

From an ABC perspective, importance sampling works much the same as rejection sampling. The
target distribution is $\pi_{ABC}(\theta,s|s_{obs})$, and the importance distribution on joint parameter value and summary statistics space is $g(\theta,s)=p(s|\theta)g(\theta)$. As a result, the (unnormalised) importance weights are computed as
\[
	\frac{\pi_{ABC}(\theta,s|s_{obs})}{g(\theta,s)} \propto \frac{K_h(\|s-s_{obs}\|)p(s|\theta)\pi(\theta)}{p(s|\theta)g(\theta)} =  \frac{K_h(\|s-s_{obs}\|)\pi(\theta)}{g(\theta)} := \tilde{w}(\theta),
\]
which is again free of intractable likelihood terms. The full ABC importance sampling algorithm is given in Algorithm 2.

\begin{table}[tbh]
\caption{\bf Algorithm 2: ABC Importance Sampling}
\noindent {\it Inputs:}
\begin{itemize}
\item A target posterior density $\pi(\theta|y_{obs})\propto p(y_{obs}|\theta)\pi(\theta)$, consisting of a prior distribution $\pi(\theta)$ and a procedure for generating data under the model  $p(y_{obs}|\theta)$.
\item An importance sampling density $g(\theta)$, with $g(\theta)>0$ if $\pi(\theta|y_{obs})>0$.
\item An integer $N>0$.
\item A kernel function $K_h(u)$ and scale parameter $h>0$.
\item A low dimensional vector of summary statistics $s=S(y)$.
\\
\end{itemize}

\noindent {\it Sampling:}\\
\noindent For $i=1, \ldots, N$:
\begin{enumerate}
\item \label{alg:importance:step1} Generate $\theta^{(i)}\sim g(\theta)$ from importance sampling density $g$.
\item Generate $y^{(i)}\sim p(y|\theta^{(i)})$ from the model. 
\item Compute summary statistic $s^{(i)}=S(y^{(i)})$.
\item Compute weight $\tilde{w}^{(i)}=K_h(\|s^{(i)}-s_{obs}\|)\pi(\theta^{(i)})/g(\theta^{(i)})$.
\\
\end{enumerate}

\noindent {\it Output:}\\
A set of weighted parameter vectors $(\theta^{(1)},\tilde{w}^{(1)}),\ldots,(\theta^{(N)},\tilde{w}^{(N)})$ $\sim$ $\pi_{ABC}(\theta|s_{obs})$.
\end{table}

As with rejection sampling, the 
choice of the (marginal) importance distribution $g(\theta)$ is crucial to the efficiency of the algorithm. 
In standard importance sampling, if $g(\theta)\propto\tilde{f}(\theta)$ then $\tilde{w}(\theta)\propto 1$. In this case, there is no variation in the importance weights, and each sample $\theta^{(i)}$ contributes equally when computing posterior expectations via (\ref{eqn:expectation}). However, if $g(\theta)$ is different to $\tilde{f}(\theta)$ then the variability in $\tilde{w}(\theta)$ from $\theta$ means that some samples $\theta^{(i)}$ will contribute more than others in this computation. In extreme cases, Monte Carlo estimates of expectations can be highly variable when they are dominated by a small number of $\theta^{(i)}$ with relatively large weights. This is known as sample degeneracy. Accordingly, for importance sampling algorithms, the focus is on reducing the variability of $\tilde{w}(\theta)$ over $\theta$.

A common measure of the degree of
sample degeneracy is the effective sample size (ESS) \shortcite{liucw98,liu01}, 
estimated as
\begin{equation}
	\label{eqn:ess}
	ESS = \left(\sum_{i=1}^{N}[W^{(i)}]^{2}\right)^{-1}, 
	\quad 1\leq ESS\leq N,
\end{equation}
which is computed using the normalised weights  $W^{(i)} = \tilde{w}^{(i)}/\sum_{j=1}^N \tilde{w}^{(j)}$.
The $ESS$ is an estimate of the effective number of equally weighted $\theta^{(i)}$ in a given weighted sample, which can be loosely interpreted as the information content.
When $g(\theta)\propto\tilde{f}(\theta)$ so that we have samples directly from $f(\theta)$,  then $W^{(i)}=1/N$ and $ESS=N$.  However, when there is severe particle degeneracy in the extreme case where  $W^{(1)}=1$ and $W^{(i)}=0$ for $i=2,\ldots,N$, then $ESS=1$.

Specifically in the ABC framework where $\tilde{w}(\theta^{(i)})=K_h(\|s^{(i)}-s_{obs}\|)\pi(\theta^{(i)})/g(\theta^{(i)})$, if $g$ is diffuse compared to the (marginal) target distribution $\pi_{ABC}(\theta|s_{obs})$, samples $\theta^{(i)}\sim g(\theta)$ from regions of  low posterior density will tend to generate summary statistics $s^{(i)}$ that are very far from the observed statistics $s_{obs}$, and this will produce low weights $w^{(i)}$ compared to samples $\theta^{(i)}$ in regions of high posterior density. This is the same as for the standard importance sampling case. However, in ABC importance sampling, an addition\textcolor{red}{al} factor is that the importance weight $\tilde{w}(\theta)$ is a function of the kernel function $K_h(\|s-s_{obs}\|)$, which contains the stochastic term $s$. This has some implications, which are also relevant for sequential Monte Carlo-based ABC samplers, discussed in Section \ref{chap5:sec:SIS}.

When $K_h$ has non-compact support, such as when $K_h(u)=\phi(u; 0, h^2)$, where $\phi(x; \mu,\sigma^2)$ denotes the Gaussian density function with mean $\mu$ and variance $\sigma^2$, the importance weight $\tilde{w}^{(i)}$ is guaranteed to be non-zero for each $i$. However the resulting importance weight can be highly variable, depending on whether $s^{(i)}$ is close to or far from $s_{obs}$. This typically produces samples $(\theta^{(i)}, \tilde{w}^{(i)})$ with low effective sample sizes.

If $K_h$ has a compact support (and this is typical in most ABC implementations), then $\tilde{w}^{(i)}=0$ is 
likely for small $h$, even when $\theta^{(i)}$ is in a high posterior density region. This means that Algorithm 2 will return many $(\theta^{(i)},\tilde{w}^{(i)})$ for which the weight is exactly zero, resulting in low effective sample sizes, and maybe even complete algorithm failure if $\tilde{w}^{(i)}=0$ for all $i=1,\ldots,N$. As a result, a common variation of Algorithm 2 is to repeat steps 1--4 for each $i$, until a non-zero weight has been generated. 
This effectively introduces a rejection sampling step within the importance sampling algorithm. This idea (c.f. \citeNP{fearnhead+p12}) can be used to improve the $ESS$ for ABC importance sampling algorithms, regardless of the choice of the kernel, by modifying step 4 in Algorithm 2 to be
\begin{table}[h]
\caption{\bf Algorithm 3: ABC Importance/Rejection Sampling}
Same as Algorithm 2, but replacing step 4 of Sampling with:

\begin{enumerate}
\item[4.] With probability $K_h(\|s^{(i)}-s_{obs}\|)/K_h(0)$ set $\tilde{w}^{(i)}=\pi(\theta^{(i)})/g(\theta^{(i)})$, else go to 1.
\\
\end{enumerate}
\end{table}

When $K_h$ has compact support, this ensures that steps 1--4 of Algorithm 2 are repeated until $K_h(\|s^{(i)}-s_{obs}\|)$ is non-zero. When $K_h$ has non-compact support, this offers some control over the variability of the weights, as only samples for which $s^{(i)}$ is reasonably close to $s_{obs}$ are likely to be accepted.

Under Algorithm 3, in the particular case of when $K_h$ is the uniform kernel on $[-h,h]$, if in addition $g(\theta)\propto\pi(\theta)$ so that the importance distribution is proportional to the prior, then $\tilde{w}^{(i)}\propto 1$ for any $i$. This results in $W^{(i)}=1/N$ and $ESS=N$ and the ABC importance sampling algorithm
effectively reduces to the ABC rejection sampling algorithm, but without the need to compute the normalising constant $M$. This setup is very common in practice as it removes the need to compute importance weights, and to worry about algorithm performance with respect to effective sample size, which is always maximised. However, in this case, algorithm performance is dominated by the number of times steps 1--4 are repeated before a sample $\theta^{(i)}$ is accepted. In general the efficiency of Algorithm 3 is a combination of the resulting effective sample size and the number of repetitions of the sampling steps 1--4.

\subsection{Importance/rejection sampler variants}

There are many variants on ABC importance and rejection samplers. A few of these are detailed below, chosen either because of their popularity, or because of their links with particular ABC samplers discussed in later Sections.

\subsubsection{Rejection control importance sampling}
\label{sec:RC}

 \shortciteN{liucw98} developed a general importance-rejection algorithm technique known as rejection control, with the aim of reducing the number of $\theta^{(i)}$ samples that are produced with very small weights in an importance sampler. This method was exploited within an ABC sequential Monte Carlo framework by \shortciteN{sisson+ft07} and \shortciteN{peters+fs12} (see Section \ref{chap5:sec:SIS}), however it may also be implemented directly within an ABC importance sampler as outlined below.
  
Suppose that a weighted sample $(\theta^{(i)},\tilde{w}^{(i)})$ is drawn from $f(\theta)$ using an importance sampling algorithm. In order to control the size of the importance weight, $\tilde{w}^{(i)}$ is compared to some pre-specified threshold value $c>0$. If $\tilde{w}^{(i)}>c$ then the weight is considered sufficiently large, and the sample $\theta^{(i)}$ is accepted. However if $\tilde{w}^{(i)}<c$ then $\theta^{(i)}$ is probabilistically rejected, with a higher rejection rate for lower $\tilde{w}^{(i)}$. In this manner, the variability of the accepted importance weights can be reduced.
In particular, each sample $\theta^{(i)}$ 
is accepted with probability
\[
	r^{(i)}=\min\left\{1, \frac{\tilde{w}^{(i)}}{c}\right\},
\]
which results in the automatic acceptance of samples for which $\tilde{w}^{(i)}>c$ and an acceptance probability of $\tilde{w}^{(i)}/c$ otherwise. This means that larger $c$ results in less variable weights, although at the price of more rejections.
The accepted samples are then draws from the modified importance sampling distribution
\[
	g^*(\theta) = M^{-1}\min\left\{1, \frac{\tilde{w}(\theta)}{c}\right\} g(\theta)
\]
where  $\tilde{w}(\theta)=\tilde{f}(\theta)/g(\theta)$, and with normalising constant $M=\int \min\{1, \tilde{w}(\theta)/c\} g(\theta) d\theta$. As a result, setting
\begin{equation}\label{eqn:rc}
\tilde{w}^{*}(\theta)=\frac{\tilde{f}(\theta)}{g^*(\theta)}
\qquad
\tilde{w}^{*(i)}=\frac{\tilde{f}(\theta^{(i)})}{g^*(\theta^{(i)})} = M \frac{\tilde{w}^{(i)}}{r^{(i)}}
\end{equation}
means that the samples $(\theta^{(i)},\tilde{w}^{*(i)})$ will be weighted samples from $f(\theta)$ but with the property that 
\begin{equation}
	\label{eqn:VarReduce}
	Var_{g^*}\left[\frac{\tilde{f}(\theta)}{g^*(\theta)}\right] \leq Var_g\left[\frac{\tilde{f}(\theta)}{g(\theta)}\right].
\end{equation}
That is, the rejection control algorithm can reduce the variance of the importance weights \cite{liu01}.
While it may be difficult to evaluate $M = E_g\left[\min\left\{1, \frac{\tilde{w}(\theta)}{c}\right\}\right]$ analytically, it may be estimated from the samples $(\theta^{(i)},\tilde{w}^{(i)})$ via
\[
	\hat{M} 
	\approx 
	\frac{1}{N} \sum_{i=1}^N \min\left\{1, \frac{\tilde{w}^{(i)}}{c}\right\}.
\]
If an estimate of $M$ is not required, its computation can be avoided for importance sampling purposes by calculating the normalised weights $W^{*(i)} = \tilde{w}^{*(i)}/\sum_{j=1}^N\tilde{w}^{*(j)}$, as the $M$ term then cancels in numerator and denominator.

As with ABC rejection sampling (Algorithm 2), the ABC implementation of rejection importance control  targets $\pi_{ABC}(\theta,s|s_{obs})$ resulting in a weight calculation of $\tilde{w}^{(i)}=K_h(\|s^{(i)}-s_{obs}\|)\pi(\theta^{(i)})/g(\theta^{(i)})$. The full algorithm is given in Algorithm 4.

Note that while Algorithm 4 requires pre-specification of the rejection threshold $c$, a suitable value may be practically difficult to determine in advance. As such, Algorithm 4 may be alternatively executed by first implementing steps 1--3 only for $i=1,\ldots,N$, and then specifying $c$ as some quantile of the resulting empirical distribution of $\tilde{w}^{(1)},\ldots,\tilde{w}^{(N)}$. Following this, Algorithm 4 may then continue implementation from step 4 onwards for each $i=1,\ldots,N$ (e.g. \shortciteNP{peters+fs12}).

\begin{table}[tbh]
\caption{\bf Algorithm 4: ABC Rejection Control Importance Sampling}

\noindent {\it Inputs:}
\begin{itemize}
\item A target posterior density $\pi(\theta|y_{obs})\propto p(y_{obs}|\theta)\pi(\theta)$, consisting of a prior distribution $\pi(\theta)$ and a procedure for generating data under the model $p(y_{obs}|\theta)$.
\item An importance sampling density $g(\theta)$, with $g(\theta)>0$ if $\pi(\theta|y_{obs})>0$.
\item An integer $N>0$.
\item A kernel function $K_h(u)$ and scale parameter $h>0$.
\item A low dimensional vector of summary statistics $s=S(y)$.
\item A rejection control threshold $c>0$.
\\
\end{itemize}

\noindent {\it Sampling:}\\
\noindent For $i=1,\ldots,N$:
\begin{enumerate}
\item \label{alg:rejcont:step1} Generate $\theta^{(i)}\sim g(\theta)$ from importance sampling density $g$.
\item Generate $y^{(i)}\sim p(y|\theta^{(i)})$ from the model and compute summary statistic $s^{(i)}=S(y^{(i)})$.
\item Compute weight $\tilde{w}^{(i)}=K_h(\|s^{(i)}-s_{obs}\|)\pi(\theta^{(i)})/g(\theta^{(i)})$.
\item Reject $\theta^{(i)}$ with probability $1-r^{(i)}=1-\min\{1,\frac{\tilde{w}^{(i)}}{c}\}$,  and go to Step \ref{alg:rejcont:step1}.
\item Otherwise, accept $\theta^{(i)}$ and set modified weight $\tilde{w}^{*(i)}=\tilde{w}^{(i)}/r^{(i)}$.\\
\end{enumerate}

\noindent {\it Output:}\\
A set of weighted parameter vectors $(\theta^{(1)},\tilde{w}^{*(1)}),\ldots,(\theta^{(N)},\tilde{w}^{*(N)})\sim\pi_{ABC}(\theta|s_{obs})$.

\end{table}

As with ABC importance/rejection sampling (Algorithm 3), when $K_h$ has compact support, rejection control will replace those samples for which the simulated and observed summary statistics are too far apart, resulting in $\tilde{w}^{(i)}=0$. More generally, however, rejection control provides much greater control over the variability of the weights regardless of $K_h$, producing more uniform weights for larger $c$. The price for this control is the  greater number of rejections induced as $c$ increases \shortcite{peters+fs12}.

\subsubsection{$k$-nearest neighbour ABC importance sampling}

While most published descriptions of importance and rejection sampling ABC algorithms follow the format given in Algorithms 1--4, in practice it is not uncommon to deviate from these and implement a slight variation. The reason for this is that Algorithms 1--4 require pre-specification of the kernel scale parameter $h>0$, without which importance weights cannot be calculated and accept/reject decisions cannot be made. In reality, as the scale of the distances $\|s^{(i)}-s_{obs}\|$ is unlikely to be known in advance, it is difficult to pre-determine a suitable value for $h$.

Algorithm 5 presents a variation on the ABC importance sampler of Algorithm 3 that avoids pre-specification of $h$. Here a large number $N'$ of $(\theta^{(i)},s^{(i)})$ pairs are generated from the importance sampling distribution $p(s|\theta)g(\theta)$. These are the only samples that will be used in the algorithm, so the computational overheads are fixed at $N'$ draws from the model, unlike Algorithm 3 in which the number of draws is random and unknown in advance. The $N$ samples for which $s^{(i)}$ is closest to $s_{obs}$ (as measured by $\|\cdot\|$) are then identified, and $h$ determined to be the smallest possible value so that only these $N$ samples have non-zero weights (assuming a kernel $K_h$ with compact support). Once $h$ is fixed, the importance weights can be calculated as before, and the $N$ samples $(\theta^{(i)},\tilde{w}^{(i)})$ with non-zero $\tilde{w}^{(i)}$ are returned as weighted samples from $\pi_{ABC}(\theta|s_{obs})$. 

\begin{table}[tbh]
\caption{\bf Algorithm 5: ABC $k$-nn Importance Sampling}
\noindent {\it Inputs:}
\begin{itemize}
\item A target posterior density $\pi(\theta|y_{obs})\propto p(y_{obs}|\theta)\pi(\theta)$, consisting of a prior distribution $\pi(\theta)$ and a procedure for generating data under the model  $p(y_{obs}|\theta)$.
\item An importance sampling density $g(\theta)$, with $g(\theta)>0$ if $\pi(\theta|y_{obs})>0$.
\item Integers $N'\gg N>0$.
\item A kernel function $K_h(u)$ with compact support.
\item A low dimensional vector of summary statistics $s=S(y)$.
\\
\end{itemize}

\noindent {\it Sampling:}\\
\noindent For $i=1, \ldots, N'$:
\begin{enumerate}
\item \label{alg:importance:step1} Generate $\theta^{(i)}\sim g(\theta)$ from importance sampling density $g$.
\item Generate $y^{(i)}\sim p(y|\theta^{(i)})$ from the model. 
\item Compute summary statistic $s^{(i)}=S(y^{(i)})$.
\\
\end{enumerate}

\noindent 
\begin{itemize}
\item Identify the $N$-nearest neighbours of $s_{obs}$ as measured by $\|s^{(i)}-s_{obs}\|$.
\item Index these nearest neighbours by $[1],\ldots,[N]$.
\item Set $h$ to be the largest possible value such that $K_h( \max_i\{\|s^{([i])}-s_{obs}\|\})=0$.
\item Compute weights $\tilde{w}^{([i])}=K_h(\|s^{([i])}-s_{obs}\|)\pi(\theta^{([i])})/g(\theta^{([i])})$ for $i=1,\ldots, N$.
\\
\end{itemize}

\noindent {\it Output:}\\
A set of weighted parameter vectors $(\theta^{([1])},\tilde{w}^{([1])}),\ldots,(\theta^{([N])},\tilde{w}^{([N])})$ $\sim$ $\pi_{ABC}(\theta|s_{obs})$.
\end{table}

This approach is explicitly used in e.g. \shortciteN{beaumont+zb02} and \shortciteN{blum+nps13}, and implicitly in many other ABC implementations. The differences between Algorithms 3 and 5 may seem small -- if the value of $h$
determined in Algorithm 5 was used in Algorithm 3, then (assuming the same pseudo-random numbers used in the appropriate places) the resulting draws from $\pi_{ABC}(\theta|s_{obs})$ would be identical. However, Algorithm 5 is based on a $k$-nearest neighbour algorithm for density estimation of the ABC likelihood function, and so possesses very different theoretical properties compared to Algorithm 3. This $k$-nearest neighbour approach is discussed and analysed in detail in the rejection sampling context by \shortciteN{biau+cg15}.

\subsubsection{ABC rejection sampling with stopping rule}

A version of ABC rejection sampling (Algorithm 1) which similarly does not require pre-specification of the kernel scale parameter $h>0$ is presented in Example 6 (Section 7.2) of \shortciteN{sisson+fb17} (this volume). We do not reproduce this algorithm here for brevity.
The algorithm identifies the smallest value of $h$ needed to accept exactly $N$ samples before some stopping rule is achieved. This stopping rule could be based on an overall computational budget (such as using exactly $N'$ total draws from $p(s|\theta)$), or on some perceived level of accuracy of the resulting ABC posterior approximation.
If the stopping rule is based on an overall computational budget of exactly $N'$ draws from $p(s|\theta)$ (and again the same pseudo-random numbers), this algorithm will produce exactly the same final samples from $\pi_{ABC}(\theta|s_{obs})$ as Algorithm 1, were the ABC rejection sampler to adopt the identified choice of $h$. Of course, the advantage here is that the value of $h$ is automatically determined.

\subsubsection{Rejection-based ABC algorithms for expensive simulators}

It is not uncommon for the data generation step $y^{(i)}\sim p(y|\theta^{(i)})$ in ABC algorithms to be expensive, and thereby dominate the computational overheads of the algorithms. While there are a few principled ways to mitigate this (see discussion of \shortciteNP{prangle+ek17} and \citeNP{everitt+r17} in Section \ref{sec:smcSamplers}), within rejection-based ABC algorithms it is sometimes possible to reject a proposed sampler $\theta^{(i)}$ {\em before} generating the data $y^{(i)}\sim p(y|\theta^{(i)})$. 
To see this, note that e.g.~step 4 of Algorithm 3
\begin{enumerate}
\item[4.] With probability $K_h(\|s^{(i)}-s_{obs}\|)/K_h(0)$ set $\tilde{w}^{(i)}=\pi(\theta^{(i)})/g(\theta^{(i)})$, else go to 1.
\end{enumerate}
can be alternatively implemented as 
\begin{enumerate}
\item[4.] With probability $\pi(\theta^{(i)})/g(\theta^{(i)})$ set $\tilde{w}^{(i)}=K_h(\|s^{(i)}-s_{obs}\|)/K_h(0)$, else go to 1.
\end{enumerate}
This means that steps 2 and 3 of Algorithm 3 (generate $y^{(i)}\sim p(y|\theta^{(i)})$ and compute $s^{(i)}=S(y^{(i)})$) need not be performed until the event in step 4 with probability $\pi(\theta^{(i)})/g(\theta^{(i)})$ has occurred.
This allows for a possible early rejection of $\theta^{(i)}$ before any data generation needs to take place. (Note that if $g(\theta)=\pi(\theta)$ there is no benefit to be gained.) This modification trades some computational savings for weights $\tilde{w}^{(i)}$ constructed from different terms, and thereby  having different variance properties.
This idea, which is a standard technique in standard sequential Monte Carlo samplers (e.g. \shortciteNP{delmoral06}), can be implemented in any rejection-based ABC algorithm, including ABC-MCMC and ABC-SMC samplers (Sections \ref{sec:mcmc} and \ref{chap5:sec:SIS}).

\subsubsection{Marginal ABC samplers}
\label{sec:pseudo-marginal}

Until now we have presented ABC algorithms as producing samples $(\theta^{(i)},s^{(i)})$ exactly from the joint distribution $\pi_{ABC}(\theta,s|s_{obs})\propto K_h(\|s-s_{obs}\|)p(s|\theta)\pi(\theta)$.
As a result, samples $\theta^{(i)}$ from the ABC approximation to the posterior $\pi(\theta|y_{obs})$ given by
\[
	\pi_{ABC}(\theta|s_{obs})\propto \int K_h(\|s-s_{obs}\|)p(s|\theta)\pi(\theta) ds
\]
may be obtained by marginalising over the realised $s^{(i)}$.
An alternative approach to construct an ABC algorithm could be to directly target the (marginal) posterior $\pi_{ABC}(\theta|s_{obs})$ rather than the joint posterior $\pi_{ABC}(\theta,s|s_{obs})$. This approach becomes apparent when noting that $\pi_{ABC}(\theta|s_{obs})$ can be estimated pointwise (up to proportionality), for fixed $\theta$, as
\[
	\int K_h(\|s-s_{obs}\|)p(s|\theta)\pi(\theta) ds\approx\frac{\pi(\theta)}{T}\sum_{t=1}^TK_h(\|s(t)-s_{obs}\|) := \hat{\pi}_{ABC}(\theta|s_{obs}),
\]
where $s(1),\ldots,s(T)\sim p(s|\theta)$ are $T$ independent draws of summary statistics from the partial likelihood $p(s|\theta)$ for a given $\theta$. This Monte Carlo estimate of $\pi_{ABC}(\theta|s_{obs})$ is unbiased up to proportionality (in that $E_{s|\theta}[\hat{\pi}_{ABC}(\theta|s_{obs})]\propto\pi_{ABC}(\theta|s_{obs})$), and so $\hat{\pi}_{ABC}(\theta|s_{obs})$ may be used in place of $\pi_{ABC}(\theta|s_{obs})$ in a standard rejection or importance sampler which targets $\pi_{ABC}(\theta|s_{obs})$. Using this substitution will produce a random, estimated acceptance probability or importance weight. However, because it is also unbiased (up to proportionality), the resulting target distribution will remain the same as if the exact weight had been used i.e. $\pi_{ABC}(\theta|s_{obs})$, although the sampler weights/acceptances will become more variable. This is the so-called marginal ABC sampler (e.g. \shortciteNP{marjoram+mpt03}, 	\citeNP{reeves+p05}, \shortciteNP{sisson+ft07}, \shortciteNP{ratmann+ahwr09}, \shortciteNP{toni+wsis09}, \shortciteNP{peters+fs12}, among others).

As with standard Monte Carlo estimates, the number of Monte Carlo draws $T$ affects the variability of the ABC posterior estimator. \shortciteN{bornn+psw17} explore the question of how many draws, $T$, produces the most efficient overall sampler in the context of ABC rejection and Markov chain Monte Carlo algorithms. If $T$ is large, the estimate of $\pi_{ABC}(\theta|s_{obs})$ is accurate and so the acceptance probability is accurate but at the cost of many Monte Carlo draws, however if $T$ is small, the acceptance probability is highly variable but is much cheaper to evaluate. When using a uniform kernel $K_h$, \shortciteN{bornn+psw17}  conclude that in fact, $T=1$ is the most efficient, as (loosely) the combination of $T$ draws used to accept one $\theta^{(i)}$ could be better used to accept up to $T$ different $\theta^{(i)}$'s, each using one Monte Carlo draw per ABC posterior estimate.

The idea of the marginal ABC sampler is closely related to the construction of the more recently developed pseudo-marginal sampler 
\cite{beaumont03,andrieu+r09}, a more general class of likelihood-free sampler that has gained popularity outside of the ABC setting. 
Here, rather than treating $\hat{\pi}_{ABC}(\theta|s_{obs})$ as an unbiased estimate of $\pi_{ABC}(\theta|s_{obs})$ in an algorithm that targets $\pi_{ABC}(\theta|s_{obs})$, an alternative joint posterior distribution can be constructed
\begin{equation}
	\label{eqn:joint1:T}
	\pi_{ABC}(\theta,s(1),\ldots,s(T)|s_{obs})\propto
	\left[\frac{1}{T}\sum_{t=1}^TK_h(\|s(t)-s_{obs}\|)\right]
	\left[\prod_{t=1}^Tp(s(t)|\theta)\right]
	\pi(\theta),
\end{equation}
which is defined over the joint posterior of $\theta$ and all $T$ summary statistic replicates (e.g. \shortciteNP{delmoral+dj12,sisson+f11}), where $T=1$ gives the usual ABC joint posterior $\pi_{ABC}(\theta,s|s_{obs})$. A useful property of this form of joint posterior is that the $\theta$-marginal distribution is the same for any value of $T$, and in particular
\[
	\int\ldots\int \pi_{ABC}(\theta,s(1),\ldots,s(T)|s_{obs}) ds(1)\ldots ds(T)
	=
	\pi_{ABC}(\theta|s_{obs}).
\]
This means that any sampler targeting $\pi_{ABC}(\theta,s(1),\ldots,s(T)|s_{obs})$ can produce samples from $\pi_{ABC}(\theta|s_{obs})$. 
Consider now an importance sampler targeting $\pi_{ABC}(\theta,s(1),\ldots,s(T)|s_{obs})$ with the importance sampling density
\[
	g(\theta,s(1),\ldots,s(T)) = g(\theta)\prod_{t=1}^Tp(s(t)|\theta).
\]
The resulting importance weight is
\begin{equation*}\begin{array}{ll}
	\frac{\pi_{ABC}(\theta,s(1),\ldots,s(T)|s_{obs})}{g(\theta,s(1),\ldots,s(T))}
	&\propto
	\frac{\left[\frac{1}{T}\sum_{t=1}^TK_h(\|s(t)-s_{obs}\|)\right]
	\left[\prod_{t=1}^Tp(s(t)|\theta)\right]
	\pi(\theta)}{g(\theta)\prod_{t=1}^Tp(s(t)|\theta)} \\
	&=
	\frac{\hat{\pi}_{ABC}(\theta|s_{obs})}{g(\theta)}.
	\end{array}
\end{equation*}
This means that any marginal ABC sampler targeting $\pi_{ABC}(\theta|s_{obs})$ through the unbiased estimate of the ABC posterior given by $\hat{\pi}_{ABC}(\theta|s_{obs})$ is directly equivalent to an exact algorithm targeting $\pi_{ABC}(\theta,s(1),\ldots,s(T)|s_{obs})$. That is, all marginal ABC samplers are justified by their equivalent joint space ABC algorithm.

This idea also extends to using unbiased approximations of posterior distributions within MCMC samplers (see next Section), where the technique has expanded beyond ABC algorithms to more general target distributions. Here it is more generally known as pseudo-marginal Monte Carlo methods.
See \shortciteN{andrieu+sv17} (this volume)
for a more detailed discussion of the connections between ABC marginal samplers and pseudo-marginal MCMC methods.

\section{Markov chain Monte Carlo methods}
\label{sec:mcmc}

Markov chain Monte Carlo (MCMC) methods are a highly accessible class of algorithms for obtaining samples from complex distributions (e.g. \shortciteNP{brooksgjm11}). By constructing a Markov chain with the target distribution of interest as its limiting distribution, following chain convergence,
a realised random sample path from this chain will behave like a (serially correlated) sample from the target distribution,. Their strong performance and simplicity of implementation has made MCMC algorithms the dominant Monte Carlo method for the past two decades \shortcite{brooksgjm11}.
As such, it is only natural that MCMC-based ABC algorithms have been developed.

\subsection{ABC MCMC samplers}

The Metropolis-Hastings algorithm is the most popular class of MCMC algorithm. Given the current chain state $\theta^{(i)}$, the next value in the sequence is obtain by sampling a candidate value $\theta'$ from a proposal distribution $\theta' \sim g(\theta^{(i)}, \theta)=g(\theta|\theta^{(i)})$, which is then accepted
with probability $a(\theta,\theta')=\min\left\{1, \frac{f(\theta')\textcolor{red}{g}(\theta', \theta^{(i)}) }{f(\theta)\textcolor{red}{g}(\theta^{(i)}, \theta')}\right\}$ so that $\theta^{(i+1)}=\theta'$, or otherwise rejected so that $\theta^{(i+1)}=\theta^{(i)}$. Under this mechanism the target distribution is $f(\theta)$, and there is great flexibility in the choice of the proposal distribution $g$. 
An implementation of this sampler in the ABC setting is given in Algorithm 6.
ABC MCMC algorithms were originally developed by  \shortciteN{marjoram+mpt03}. See e.g.   \shortciteN{bortot+cs07}, \shortciteN{wegmann+le09}, \shortciteN{ratmann+ahwr09}, \shortciteN{sisson+f11} and \shortciteN{andrieu+sv17} (this volume)
for more discussion on ABC MCMC samplers.

As with ABC importance and rejection samplers, the target distribution of ABC MCMC algorithms is the joint ABC posterior $\pi_{ABC}(\theta,s|s_{obs})$. On this space the proposal distribution becomes
\[
	g[(\theta,s),(\theta',s')] =g(\theta,\theta') p(s'|\theta'),
\]
and as a result the acceptance probability of the proposed move from $(\theta^{(i)}, s^{(i)})$ to $(\theta', s')\sim g[(\theta^{(i)},s^{(i)}),(\theta',s')]$ becomes
$a[(\theta^{(i)},s^{(i)}),(\theta',s')] = \min\{1,\alpha[(\theta^{(i)},s^{(i)}),(\theta',s')]\}$, where
\begin{eqnarray*}
	\alpha[(\theta^{(i)},s^{(i)}),(\theta',s')]
	&=&
	\frac{\pi_{ABC}(\theta',s'|s_{obs})g[(\theta',s'),(\theta^{(i)},s^{(i)})]}{\pi_{ABC}(\theta^{(i)},s^{(i)}|s_{obs})g[(\theta^{(i)},s^{(i)}),(\theta',s')]}\\
	&=&
	\frac{K_h(\|s'-s_{obs}\|)p(s'|\theta')\pi(\theta')}{K_h(\|s^{(i)}-s_{obs}\|)p(s^{(i)}|\theta^{(i)})\pi(\theta^{(i)})}\frac{g(\theta',\theta^{(i)}) p(s^{(i)}|\theta^{(i)})}{g(\theta^{(i)},\theta') p(s'|\theta')}\\
	&=&
	\frac{K_h(\|s'-s_{obs}\|)\pi(\theta')}{K_h(\|s^{(i)}-s_{obs}\|)\pi(\theta^{(i)})}\frac{g(\theta',\theta^{(i)})}{g(\theta^{(i)},\theta')},
\end{eqnarray*}
which is free of intractable likelihood terms, $p(s|\theta)$, and so may be directly evaluated.

\begin{table}[tbh]
\caption{\bf Algorithm 6: ABC Markov Chain Monte Carlo Algorithm}
\noindent {\it Inputs:}
\begin{itemize}
\item A target posterior density $\pi(\theta|y_{obs})\propto p(y_{obs}|\theta)\pi(\theta)$, consisting of a prior distribution $\pi(\theta)$ and a procedure for generating data under the model $p(y_{obs}|\theta)$.
\item A Markov proposal density $g(\theta, \theta')=g(\theta'|\theta)$. 
\item An integer $N>0$.
\item A kernel function $K_h(u)$ and scale parameter $h>0$.
\item A low dimensional vector of summary statistics $s=S(y)$.
\\
\end{itemize}

\noindent {\it Initialise:}\\
Repeat:
\begin{enumerate}
\item Choose an initial parameter vector $\theta^{(0)}$ from the support of $\pi(\theta)$.
\item Generate  $y^{(0)}\sim p(y|\theta^{(0)})$ from the model and compute summary statistics $s^{(0)}=S(y^{(0)})$.
\end{enumerate}
until $K_h(\|s^{(0)}-s_{obs}\|)>0$.
\\

\noindent {\it Sampling:}\\
\noindent For $i=1, \ldots, N$:
\begin{enumerate}
\item Generate candidate vector $\theta'\sim g(\theta^{(i-1)}, \theta)$ from the proposal density $g$
\item Generate $y'\sim p(y|\theta')$ from the model and compute summary statistics $s'=S(y')$.
\item With probability 
\[
	\min\left\{1, \frac{K_h(\|s'-s_{obs}\|)\pi(\theta')g(\theta', \theta^{(i-1)})}{K_h(\|s^{(i-1)}-s_{obs}\|)\pi(\theta^{(i-1)})g(\theta^{(i-1)}, \theta')} 
	\right\}
\]
set $(\theta^{(i)},s^{(i)})=(\theta',s')$. Otherwise set $(\theta^{(i)},s^{(i)})=(\theta^{(i-1)}, s^{(i-1)})$. \\
\end{enumerate}

\noindent {\it Output:}\\
A set of correlated parameter vectors $\theta^{(1)},\ldots,\theta^{(N)}$ from a Markov chain with stationary distribution
$\pi_{ABC}(\theta | s_{obs})$.

\end{table}

Algorithm 6 satisfies the detailed balance (time reversibility) condition with respect to $\pi_{ABC}(\theta,s|s_{obs})$, which ensures that $\pi_{ABC}(\theta,s|s_{obs})$ is the stationary distribution of the Markov chain. Detailed balance states that
\begin{equation*}
	\pi_{ABC}(\theta, s|s_{obs})P[(\theta, s),(\theta', s')]  = \pi_{ABC}(\theta',s'|s_{obs})P[(\theta', s'),(\theta, s)],
\end{equation*}
where
the Metropolis-Hastings transition kernel $P$ is given by
\[
	P[(\theta, s),(\theta', s')]=g[(\theta, s),(\theta', s')]a[(\theta, s),(\theta', s')].
\]
Assuming that (without loss of generality) $a[(\theta', s'),(\theta, s)]=\min\{1,\alpha[(\theta',s'),(\theta,s)]\}=1$ (and so $a[(\theta, s),(\theta', s')]=\alpha[(\theta,s),(\theta',s')]$), the
detailed balance condition  is satisfied since
\begin{eqnarray*}
&&\pi_{ABC}(\theta, s|s_{obs})P[(\theta, s),(\theta', s')]\\
& =& \pi_{ABC}(\theta, s|s_{obs})g[(\theta, s),(\theta', s')]\alpha[(\theta, s),(\theta', s')]\\
&=&\frac{K_h(\|s-s_{obs}||)p(s|\theta)\pi(\theta)}{z}g(\theta,\theta')p(s'|\theta')\frac{K_h(\|s'-s_{obs}\|)\pi(\theta')g(\theta',\theta)}{K_h(\|s-s_{obs}\|)\pi(\theta)g(\theta,\theta')}\\
&=&\frac{K_h(\|s'-s_{obs}\|)p(s'|\theta')\pi(\theta')}{z}g(\theta',\theta)p(s|\theta)\\
&=&\pi_{ABC}(\theta',s'|s_{obs})P[(\theta', s'),(\theta, s)],
\end{eqnarray*}
where $z=\int\int K_h(\|s-s_{obs}\|)p(s|\theta)\pi(\theta)dsd\theta$ is the normalisation constant of $\pi_{ABC}(\theta,s|s_{obs})$ (e.g. \shortciteNP{sisson+f11}).

\shortciteN{marjoram+mpt03} found that the ABC MCMC algorithm offered an improved acceptance rate over rejection sampling-based ABC algorithms with the same scale parameter $h$, although at the price of serial correlation in the Markov chain sample path $\theta^{(1)},\ldots,\theta^{(N)}$.
Thus, for kernels $K_h$ with compact support, the same mechanism that causes many rejections or zero weights in ABC rejection and importance samplers, now results in many rejected proposals in the ABC MCMC algorithm. The difference here is that the chain simply remains at the current state $\theta^{(i)}$ for long periods of time, giving additional posterior weight to $\theta^{(i)}$.
Techniques for improving the performance of standard MCMC algorithms may also be applied to ABC MCMC samplers. However there is one feature of ABC MCMC that is different to that of the standard algorithm, that is particularly acute when using kernel functions $K_h$ with compact support.

Consider a proposed move from $\theta^{(i)}$ to $\theta'$. In standard MCMC, the acceptance probability is based on the relative density of the posterior evaluated at $\theta'$ compared to that evaluated at $\theta^{(i)}$. In ABC MCMC the density of the posterior at $\theta'$ is determined through the ability of the model to generate a summary statistic $s'\sim p(s|\theta')$ that is close to $s_{obs}$ as measured through $K_h(\|s'-s_{obs}\|)$. That is, to move to $\theta'$, a summary statistic $s'$ must be generated that is close enough to $s_{obs}$. This is the standard ABC mechanism. However, the result of this for the ABC MCMC algorithm is that it means that the acceptance rate of the sampler is directly related to the value of the (intractable) likelihood function evaluated at $\theta'$. As a result, the sampler may mix rapidly in regions of high posterior density, but will have much worse mixing in regions of relatively low posterior density \shortcite{sisson+ft07}. For this reason, ABC-MCMC samplers can often get stuck in regions of low posterior density for long periods time, effectively producing convergence issues for the algorithm. 

This effect is more pronounced when the kernel $K_h$ has compact support, such as the uniform kernel on $[-h,h]$ which is endemic in ABC implementations, although it is still present for kernels defined on the real line, such as the Gaussian density kernel. 
In a study of  `sojourn time' within ABC MCMC samplers (that is, the number of consecutive iterations in the sampler in which a univariate parameter $\theta$ remained above some high threshold), \citeN{sisson+f11} found empirically that samplers with uniform kernels had a substantially higher expected sojourn time than samplers with Gaussian kernels, indicating that the latter had superior chain mixing in distributional tails. Despite this, ABC MCMC samplers are routinely implemented with uniform kernels $K_h$.

Chain mixing can be improved by alternatively targeting the joint posterior distribution $\pi_{ABC}(\theta,s(1),\ldots,s(T)|s_{obs})$ given by (\ref{eqn:joint1:T}). Under this (pseudo) marginal sampler framework (Section \ref{sec:pseudo-marginal}) as $T\rightarrow\infty$, the mixing properties of the ABC MCMC  approach that of the equivalent standard MCMC sampler directly targeting $\pi_{ABC}(\theta|s_{obs})$ (if it would be possible to numerically evaluate the density function). \citeN{sisson+f11} empirically demonstrated this improvement, as measured in sojourn times, as $T$ increases. Of course, this improvement of chain mixing comes at the price of overall sampler performance as the computational overheads of generating $s(1),\ldots,s(T)$ for large $T$ would be extremely high. The results of \shortciteN{bornn+psw17}, that $T=1$ is the optimum efficiency choice for uniform kernels $K_h$, also hold for ABC MCMC samplers.

\subsection{Augmented space ABC-MCMC samplers}
\label{sec:epsaug}

The standard ABC MCMC sampler (Algorithm 6) requires pre-specification of the kernel scale parameter $h$. As with ABC rejection and importance samplers, there are a number of ways in which lack of knowledge of a suitable value for the kernel scale parameter can be incorporated into the basic algorithm. Most of these methods also attempt to improve chain mixing over the standard algorithm which uses a fixed, low value of $h$.
At the very simplest level,  this could involve adaptively adjusting $h$ as a function of $\|s-s_{obs}\|$ at the current and proposed  states of the chain, and either allow $h$ to slowly reduce to some target value to improve convergence at the start of the sampler (e.g. \shortciteNP{ratmann+jhsrw07}, \shortciteNP{sisson+f11}, p.325), or adaptively choose $h$ to achieve some pre-determined overall sampler acceptance probability.

Augmenting the dimension of the target distribution is a common strategy to improve the performance of Monte Carlo algorithms.
In order to help the ABC MCMC sampler escape from regions of low posterior density,  \shortciteN{bortot+cs07} proposed augmenting the joint ABC posterior $\pi_{ABC}(\theta,s|s_{obs})$ to additionally include the kernel bandwidth $h$,  treating this as an unknown additional parameter. The resulting joint posterior distribution is given by
\begin{equation*}
	\pi_{ABC}(\theta, s, h|s_{obs})\propto K_h(\|s-s_{obs}\|)p(s|\theta)\pi(\theta)\pi(h),
\end{equation*}
and the resulting ABC approximation to the partial posterior $\pi(\theta|s_{obs})$ is then given by
\begin{equation}
\label{eqn:aug-approx}
	\dot{\pi}_{ABC}(\theta|s_{obs})=\int \int\pi_{ABC}(\theta,s,h|s_{obs})dsdh
\end{equation}
where $h>0$. 
Here $h$ is treated as a tempering parameter in the manner of simulated tempering \cite{geyer+t95}, with larger and smaller values respectively corresponding to ``hot'' and ``cold'' tempered posterior distributions. 
Larger values of $h$ increase the scale of the kernel density function $K_h$, under which the sampler is more likely to accept proposed moves and thereby alleviating the sampler's mixing problems, although at the price of a less accurate posterior approximation. Lower values of $h$ produce a more accurate posterior approximation, but will induce slower chain mixing.
The density $\pi(h)$ is a pseudo-prior, which serves to influence the mixing of the sampler through the tempered distributions.

Note that the augmented space ABC posterior approximation $\dot{\pi}_{ABC}(\theta|s_{obs})$ given by (\ref{eqn:aug-approx}) will in general be different to that of $\pi_{ABC}(\theta|s_{obs})$ as the latter contains a fixed value of $h$, whereas the former integrates over the uncertainty inherent in this parameter. 
Rather than use (\ref{eqn:aug-approx}) as the final ABC approximation to $\pi(\theta|s_{obs})$, \shortciteN{bortot+cs07} chose to remove those samples $(\theta^{(i)},s^{(i)},h^{(i)})$ for which $h^{(i)}$ was considered too large to come from a good approximation to $\pi(\theta|s_{obs})$. In particular, they examined the distribution of $\theta^{(i)}|h^{(i)}\leq h^*$, aiming to choose the largest value of $h^*$ such that the distribution of $\theta^{(i)}|h^{(i)}\leq h^*$ did not change if $h^*$ was reduced further. The resulting ABC posterior approximation is therefore given by
\begin{equation*}
	\ddot{\pi}_{ABC}(\theta|s_{obs})=\int_0^{h^*} \int\pi_{ABC}(\theta,s,h|s_{obs})dsdh.
\end{equation*}
This approach effectively permits an {\it a posteriori} evaluation of an appropriate value $h^*$ such that  the approximation $\ddot{\pi}_{ABC}(\theta|s_{obs})$ is as close as possible (subject to Monte Carlo variability) to the true posterior $\pi(\theta|s_{obs})$.

A similar idea was explored by \shortciteN{baragatti+gp13} in an ABC version of the parallel tempering algorithm of \citeN{geyer+t95}. Here $M>1$ parallel ABC MCMC chains are implemented with different kernel density scale parameters $h_M<h_{M-1}<\ldots<h_1$, with state transitions allowed between chains so that the states of the more rapidy mixing chains (with higher $h$ values) can propagate down to the more slowly mixing chains (with lower $h$). The final ABC posterior approximation is the output from the chain with $h=h_M$.
A related augmented space ABC sampler based on the equi-energy MCMC sampler of \shortciteN{kou+zw06} could similarly be implemented.

\shortciteN{ratmann+ahwr09} take the auxiliary space ABC sampler of \shortciteN{bortot+cs07} beyond the solely mechanical question of improving Markov chain mixing, and towards estimation of the distribution of $s-s_{obs}$ under the model. This is more in line with the ABC approximation $\dot{\pi}_{ABC}(\theta|s_{obs})$ given by (\ref{eqn:aug-approx}), and the interpretation of the ABC approximation to $\pi(\theta|s_{obs})$ as an exact model in the presence of model error due to \citeN{wilkinson13}. It additionally allows an assessment of model adequacy. Instead of comparing $s$ to $s_{obs}$ through $K_h(\|s-s_{obs}\|)$ with a single $h$, \shortciteN{ratmann+ahwr09} alternatively make the comparison independently and univariately for each of the $q$ summary statistics in $s=(s_1,\ldots,s_q)^\top$  via $K_{h_r}(\tau_r-|s_r-s_{obs,r}|)$ for $r=1,\ldots,q$. Here, $\tau_r$ is the parameter denoting the true but unknown discrepancy between the $r$-th summary statistics of $s$ and $s_{obs}$, i.e. $|s_r-s_{obs,r}|$, and so if $\tau_r=0$ then the model can adequately explain the observed data as described through the $r$-th summary statistic. The full model has a joint target distribution of
\begin{eqnarray*}
	&&\pi_{ABC}(\theta,s(1),\ldots,s(T),\tau|s_{obs}) \\
	&\propto&
	\min_{r} \left[\frac{1}{Th_{r}}\sum_{t=1}^T K_{h_{r}}\left(\tau_r-|s_r(t)-s_{obs,r}|\right)\right]
	\left[\prod_{t=1}^T p(s(t)|\theta)\right]
	\pi(\theta)\pi(\tau),
\end{eqnarray*}
based on $T$ samples $s(1),\ldots,s(T)\sim p(s|\theta)$, where $s_r(t)$ is the $r$-th element of $s(t)$, and $\pi(\tau)=\prod_{r=1}^q\pi(\tau_r)$. 
The minimum over the univariate density estimates aims to focus the model on the most conservative estimate of model adequacy, while also reducing computation over $\tau$ to its univariate margins.
Here interest is in the posterior distribution of $\tau$ in order to determine model adequacy (i.e. if the posterior marginal distribution of $\pi_{ABC}(\tau_r|s_{obs})$ is centered on 0), whereas the margin specific kernel scale parameters $h_r$ are determined via standard kernel density estimation arguments over the observed sample $|s_r(t)-s_{obs,r}|$ for $t=1,\ldots,T$.

\subsection{Other ABC MCMC samplers}
\label{sec:alternative}

The field of MCMC research with tractable target distributions is fairly mature, and it is not difficult to imagine that many known techniques can be directly applied to ABC MCMC algorithms to improve their performance. Different forms of algorithms include 
Hamiltonian Monte Carlo ABC samplers \shortcite{meeds+lw15} which use a moderate number of simulations under the intractable model to produce an ABC estimate of the otherwise intractable gradient of the potential energy function,
multi-try Metropolis ABC (\shortciteNP{aandahlThesis}, \shortciteNP{kobayashi+k15}) which uses multiple proposals to choose from at each stage of the sampler to ensure improved mixing and acceptance rates,
in addition to the various augmented space samplers discussed in the previous Section \shortcite{bortot+cs07,ratmann+ahwr09,baragatti+gp13}. 
Of course, transdimensional ABC MCMC samplers can also be implemented for multi-model posterior inference. 

General improvements in efficiency can be obtained by using quasi Monte Carlo ABC methods to form efficient proposal distributions \shortcite{cabras+nr15}.
In a similar manner, \citeN{neal12} developed a coupled ABC MCMC sampler which uses the same random numbers to generate the summary statistics for different parameter values, and showed this algorithm to be more efficient than the standard ABC MCMC sampler.

Within the standard ABC MCMC sampler, \shortciteN{kousathanas+lhqfw16} proposed using a subset $\tilde{s}\subseteq s$ of the vector of summary statistics within the acceptance probability when updating a subset of the model parameters conditional on the rest. Here the idea was to reduce the dimension of the comparison $\|\tilde{s}-\tilde{s}_{obs}\|$ within the kernel $K_h$ to increase the efficiency and mixing of the algorithm. \citeN{rodrigues17} developed a related algorithm based on the Gibbs sampler.

\citeN{lee+l14} present an analysis of the variance bounding and geometric ergodicity properties of three reversible kernels used for ABC MCMC, previously suggested by \shortciteN{lee+ad12}, which are based on the uniform kernel $K_h$.
Given that current state of the chain is $\theta^{(i)}$ and a proposed new state is drawn from $\theta'\sim g(\theta^{(i)},\theta)$, the following algorithms were examined (where $I(\cdot)$ denotes the indicator function):
\begin{itemize}
\item {\em Method 1:}
Draw $s'(1),\ldots,s'(T)\sim p(s|\theta')$.

Accept the move $\theta^{(i+1)}=\theta'$ (and $s(t)=s'(t)$ $\forall t$) with probability
\[
	\min\left\{1,\frac{\left[\sum_{t=1}^{T} I(\|s'(\textcolor{red}{t})-s_{obs}\|\leq h)\right]\pi(\theta')g(\theta',\theta^{(i)})}
	{\left[\sum_{t=1}^{T}I(\|s(t)-s_{obs}\|\leq h)\right]\pi(\theta^{(i)})g(\theta^{(i)},\theta')}\right\}
\]
else reject and set $\theta^{(i+1)}=\theta^{(i)}$.
\item {\em Method 2:}
Draw $s(1),\ldots,s(T-1)\sim p(s|\theta^{(i)})$ and $s'(1),\ldots,s'(T)\sim p(s|\theta')$.

Accept the move $\theta^{(i+1)}=\theta'$ with probability
\[
	\min\left\{1,\frac{\left[\sum_{t=1}^{T} I(\|s'(\textcolor{red}{t})-s_{obs}\|\leq h)\right]\pi(\theta')g(\theta',\theta^{(i)})}
	{\left[1+\sum_{t=1}^{T-1}I(\|s(t)-s_{obs}\|\leq h)\right]\pi(\theta^{(i)})g(\theta^{(i)},\theta')}\right\}
\]
else reject and set $\theta^{(i+1)}=\theta^{(i)}$.
\item {\em Method 3:}
Reject the move and set $\theta^{(i+1)}=\theta^{(i)}$ with probability
\[
	1-\min\left\{1,\frac{\pi(\theta')g(\theta',\theta^{(i)})}{\pi(\theta^{(i)})g(\theta^{(i)},\theta)}\right\}.
\]
For $T=1,2,\ldots$ 
draw $s(T)\sim p(s|\theta^{(i)})$ and $s'(T)\sim p(s|\theta')$
until $\sum_{t=1}^TI(\|s(t)-s_{obs}\|\leq h)+I(\|s'(t)-s_{obs}\|\leq h)\geq 1$.

If $I(\|s'(T)-s_{obs}\|\leq h)=1$ then set $\theta^{(i+1)}=\theta'$ else set $\theta^{(i+1)}=\theta^{(i)}$.
\end{itemize}

Method 1 is the acceptance probability constructed from the standard Monte Carlo estimate of the ABC posterior $\hat{\pi}_{ABC}(\theta|s_{obs})$ using a fixed number, $T$, of summary statistic draws, as described in Section  \ref{sec:pseudo-marginal}. Method 2 is the same as Method 1, except that $T-1$ of the summary statistics of the current chain state $s|\theta^{(i)}$ are regenerated anew in the denominator of the acceptance probability. The idea here is to help the Markov chain escape regions of low posterior probability more easily than under Method 1, at the cost of higher computation. Method 3 produces a random number of summary statistic generations, with computation increasing until either $s|\theta^{(i)}$ or $s'|\theta'$ is sufficiently close to $s_{obs}$.

Under some technical conditions, \citeN{lee+l14} conclude that Methods 1 and 2 cannot be variance bounding, and that Method 3 (as with the standard Metropolis-Hastings algorithm if it were analytically tractable) can be both variance bounding and geometrically ergodic. Overall these results, 
in addition to other methods for constructing estimates of intractable likelihoods (e.g.  \shortciteNP{buchholz+c17}),
are very interesting from the perspective of future simulation-based algorithm design.

\section{Sequential Monte Carlo sampling}
\label{chap5:sec:SIS}

It can be difficult to design an importance sampling density $g(\theta)$ that is able to efficiently place a large number of samples in regions of high posterior density. Sequential Monte Carlo (SMC) and sequential importance sampling (SIS) algorithms are designed to overcome this difficulty by constructing a sequence of slowly changing intermediary distributions $f_m(\theta)$, $m=0,\ldots,M$, where $f_0(\theta)=g(\theta)$ is the initial importance sampling distribution, and $f_M(\theta)=f(\theta)$ is the target distribution of interest. A population of particles (i.e samples $\theta^{(i)}$, $i=1,\ldots,N$) is then propagated between these distributions, in sequence, so that $f_1(\theta),\ldots,f_{M-1}(\theta)$ act as an efficient importance sampling bridge between $g(\theta)$ and $f(\theta)$. There are a number of techniques available for specification of the intermediary distributions (e.g. \shortciteNP{geyer+t95}, \shortciteNP{delmoral06}). 
There is a rich literature on the construction of efficient SMC and SIS algorithms. See e.g. \shortciteN{liucw98}, \citeN{gilksb01}, \citeN{neal01}, \shortciteN{doucetfg01} and \citeN{chopin02} among others. These algorithms invariably involve some combination of three main ideas.

Given a weighted sample $(\theta^{(1)}_{m-1},w^{(1)}_{m-1}),\ldots,(\theta^{(N)}_{m-1},w^{(N)}_{m-1})$ from intermediary distribution $f_{m-1}(\theta)$, the {\em reweighting} step propagates the particles to the next intermediary distribution $f_{m}(\theta)$. This could involve a simple importance reweighting, or something more involved if hybrid importance/rejection schemes are employed (e.g. \shortciteNP{liucw98}).

Depending on the efficiency of the transitions between $f_{m-1}(\theta)$ and $f_m(\theta)$, the variability of the importance weights $w_m^{(i)}$ could be very high, with some particles having very small weights, and others having very large weights -- commonly known as particle degeneracy. This can be measured through the effective sample size (\ref{eqn:ess}) (\shortciteNP{liucw98}, \citeNP{liu01}). The {\em resampling} step is designed to replenish the particle population by resampling the particles from their empirical distribution $(\theta^{(1)}_{m},w^{(1)}_{m}),\ldots,(\theta^{(N)}_{m},w^{(N)}_{m})$. In this manner, particles with low weights in regions of low density will likely be discarded in favour of particles with higher weights in regions of higher density. Following resampling, the effective sample size will be reset to $N$ as each weight will then be set to $w^{(i)}_m=1/N$. Resampling should not occur too frequently. A common criterion is to resample when the effective sample size falls below a pre-specified threshold, typically $E=N/2$. See e.g. \shortciteN{douccm05} for a review and comparison of various resampling methods. 

Finally, the {\em move} step aims to both move the particles to regions of high probability, and increase the particle diversity in the population. The latter is important since, particularly after resampling, particles with high weights can be replicated in the sample. Any transition kernel $F_m(\theta,\theta')$ can be used for the move step, although an MCMC kernel is a common choice (e.g. \shortciteNP{gilksb01}) as it results in the importance weight being unchanged, although there is also the chance that the proposed move is rejected. Other kernels, such as $F_m(\theta,\theta')=\phi(\theta'; \theta,\sigma_m^{\textcolor{red}{2}})$ to add a random normal scatter to the particles, will require the importance weights to be modified. See e.g. \shortciteN{delmoral06} for discussion on different forms of \textcolor{red}{the} move kernel.

\subsection{Sequential importance sampling}

In the ABC framework a natural choice for the sequence of intermediary distributions is
\[
	f_m(\theta) =\pi_{ABC,h_m}(\theta,s|s_{obs})\propto K_{h_m}(\|s-s_{obs}\|)p(s|\theta)\pi(\theta),
\]
for $m=0,\ldots,M$, indexed by the kernel scale parameter, where the sequence $h_0\geq h_1\geq \ldots\geq h_M$ is a monotonic decreasing sequence. Accordingly, each successive distribution with decreasing $h_m$, will be less diffuse and a closer approximation to $\pi(\theta|s_{obs})$ \shortcite{sisson+ft07}.
A sequential importance sampling version of the ABC rejection control importance sampler (Algorithm 4) is given in Algorithm 7.

This algorithm is a particular version of the sampler proposed by \shortciteN{peters+fs12} (see also \shortciteNP{sisson+ft07}) who incorporated the partial rejection control mechanism of \shortciteN{liucw98} and \citeN{liu01} into the SMC sampler framework. When applied in the ABC setting, rejection control provides one means of controlling the otherwise highly variable particle weights. 
As with the ABC rejection control importance sampler (Algorithm 4), samples from an importance sampling distribution $g_m(\theta)$, constructed from the samples from the previous population targeting $f_{m-1}(\theta)$, are combined with the rejection control mechanism in order to target $f_m(\theta)$.


The initial sampling distribution $g(\theta)$ can be any importance sampling density, as with standard importance sampling algorithms.  There are a number of adaptive ways to construct the subsequent importance distributions $g_m(\theta)$ for $m=1,\ldots,M$, based on the population of samples from the previous intermediary distribution $(\theta^{(1)}_{m-1},w^{*(1)}_{m-1}),\ldots,(\theta^{(N)}_{m-1},w^{*(N)}_{m-1})$.
The simplest of these is to  specify $g_m(\theta)$ as some standard parametric family, such as the multivariate Normal distribution, with parameters estimated from the previous particle population (e.g. \citeNP{chopin02}).
Another option is to construct a kernel density estimate of the distribution of the previous particle population
$g_m(\theta) = \sum_{i=1}^N W^{*(i)}_{m-1}F_m(\theta^{(i)}_{m-1},\theta)$ where $W_{m-1}^{*(i)}=w_{m-1}^{*(i)}/\sum_{j=1}^Nw_{m-1}^{*(j)}$, and $F_m(\theta,\theta')$ is some forward mutation kernel describing the probability of moving from $\theta$ to $\theta'$, such as $F_m(\theta,\theta') = \phi(\theta'; \theta,\Sigma_m)$, the multivariate normal density function centred at $\theta$ and with covariance matrix $\Sigma_m$ \shortcite{delmoral06,beaumont+cmr09,peters+fs12}. 
These and other possibilities may also be constructed by first reweighting the draws from the previous population $f_{m-1}(\theta)$ so that they target $f_m(\theta)$.

If the kernel $K_h$ has compact support then step \ref{alg7:step3} of Algorithm 7 will automatically reject any $\theta^{(i)}_m$ for which $K_{h_m}(\|s^{(i)}_m-s_{obs}\|)=0$. (This also happens for Algorithm 4.) This practical outcome occurs for most ABC SIS and SMC algorithms used in practice, as use of the uniform kernel is predominant (e.g. \shortciteNP{sisson+ft07}, \shortciteNP{toni+wsis09}, \shortciteNP{beaumont+cmr09}, \shortciteNP{delmoral+dj12}), although the rejection of $\theta^{(i)}$ is sometimes hard coded as in Algorithm 3, rather than being part of a more sophisticated importance weight variance control mechanism, such as rejection control.

In the limit as rejection thresholds $c_m\rightarrow 0$ for $m=1,\ldots,M$ (and defining $0/0:=1$), the rejection control mechanism will allow all particles to proceed to the next stage of the algorithm. Therefore $c_m\rightarrow0$ represents a standard sequential importance sampler that will likely result in the collapse of the particle population (i.e. all weights $w^{*(i)}_m=0$) in the ABC setting, for low $h_m$.
However, non-zero rejection control thresholds $c_m$ permit a finer scale control over the importance weights $w^{(i)}_m$ beyond distinguishing between zero and non-zero weights, with larger $c_m$ resulting in more similar weights with less variability, though at the price of higher computation through more rejections. In this manner, rejection control provides one way in which ABC SMC algorithms may be implemented with kernels $K_h$ that are non-uniform, or have non-compact support, without which the effective sample size of the sampler would deteriorate almost immediately for low $h_m$ \shortcite{peters+fs12}.

\begin{table}[tbh]
\caption{\bf Algorithm 7: ABC Sequential Rejection Control Importance Sampling Algorithm}
\noindent {\it Inputs:}
\begin{itemize}
\item A target posterior density $\pi(\theta|y_{obs})\propto p(y_{obs}|\theta)\pi(\theta)$, consisting of a prior distribution $\pi(\theta)$ and a procedure for generating data under the model $p(y_{obs}|\theta)$.
\item A kernel function $K_h(u)$ and a sequence of scale parameters $h_0\geq h_1\geq \ldots\geq h_M$.
\item An initial sampling distribution $g(\theta)$, and a method of constructing subsequent sampling distributions $g_m(\theta)$, $m=1,\ldots,M$.
\item An integer $N>0$. 
\item A sequence of rejection control thresholds values $c_m$, $m=1,\ldots,M$.
\item A low dimensional vector of summary statistics $s=S(y)$.\\
\end{itemize}

\noindent {\it Initialise:}\\
For $i=1,\ldots,N$: 
\begin{itemize}
\item Generate $\theta^{(i)}_0 \sim g(\theta)$ from initial sampling distribution $g$.
\item Generate $y^{(i)}_0 \sim p(y|\theta^{(i)}_0)$ and compute summary statistics $s^{(i)}_0=S(y^{(i)}_0)$.
\item Compute weights 
$w^{(i)}_0=K_{h_0}(\|s^{(i)}_0-s_{obs}\|)\pi(\theta^{(i)}_0)/g(\theta^{(i)}_0)$.\\
\end{itemize}

\noindent {\it Sampling:}\\
For $m=1,\ldots,M$: 
\begin{enumerate}
\item Construct sampling distribution $g_{m}(\theta)$.
\item For $i=1,\ldots,N$:
\begin{enumerate}
\item \label{alg7:step1} Generate $\theta_m^{(i)}\sim g_m(\theta)$, $y^{(i)}_m\sim p(y|\theta^{(i)}_m)$ and compute $s^{(i)}_m=S(y^{(i)}_m)$.
\item \label{alg7:step2} Compute weight $w^{(i)}_m=K_{h_m}(\|s^{(i)}_m-s_{obs}\|)\pi(\theta^{(i)}_m)/g_m(\theta^{(i)}_m)$.
\item \label{alg7:step3} Reject $\theta^{(i)}_m$ with probability $1-r^{(i)}_m=1-\min\{1,\frac{w^{(i)}_m}{c_m}\}$, and go to step \ref{alg7:step1}.
\item Otherwise, accept $\theta^{(i)}_m$ and set modified weight $w^{*(i)}_m=w^{(i)}_m/r^{(i)}_m$.\\
\end{enumerate}
\end{enumerate}

\noindent {\it Output:}\\
A set of weighted parameter vectors $(\theta^{(1)}_M,w^{*(1)}_M),\ldots,(\theta^{(N)}_M,w^{*(N)}_M)$ drawn from $\pi_{ABC}(\theta|s_{obs})\propto \int K_{h_M}(\|s-s_{obs}\|)p(s|\theta)\pi(\theta)ds$.
\end{table}

As with the ABC rejection control importance sampler (Algorithm 4), suitable rejection thresholds may be dynamically determined during algorithm run-time by, for each $m$, first implementing steps \ref{alg7:step1} and \ref{alg7:step2} for $i=1,\ldots,N$, specifying $c_m$ as some function (such as a quantile) of the empirical distribution of the realised $w^{(1)}_m,\ldots,w^{(N)}_m$, and then continuing Algorithm 7 from step \ref{alg7:step3} onwards for each $i=1,\ldots,N$ \shortcite{peters+fs12}.

The sequence of scale parameters $h_0\geq h_1\geq\ldots\geq h_M$ in Algorithm 7 has been presented as requiring  pre-specification in order to implement the sampler. However, as with any annealing-type algorithm, identifying an efficient sequence is a challenging problem. Fortunately, as with the automatic determination of the rejection control thresholds $c_m$, choice of the scale parameters can also be automated, and one such method to achieve this is discussed in the next Section. To initialise the algorithm efficiently, setting $h_0=\infty$ would result in all particles $\theta_0^{(1)},\ldots,\theta_0^{(N)}$ having relatively similar weights $w_0^{(i)}=\pi(\theta^{(i)}_0)/g(\theta^{(i)}_0)$, as a function of the prior and initial sampling distributions.

\subsection{Sequential Monte Carlo samplers}
\label{sec:smcSamplers}

An alternative representation of population based algorithms is the sequential Monte Carlo sampler \shortcite{delmoral06}. Here, the particles are defined on the space of the path that each particle will take through the sequence of distributions $f_0(\theta),\ldots,f_M(\theta)$.  Hence, if $\theta^{(i)}_m\in\Theta$, then the path of particle $i$ through the first $m$ distributions is given by $\theta_{1:m}^{(i)} = (\theta_1^{(i)},\ldots,\theta_m^{(i)})\in\Theta^m$ for $m=1,\ldots,M$. SMC samplers explicitly implement each of the reweighting, resampling and move steps, and at their most general level have sophisticated implementations (e.g. \shortciteNP{delmoral06}). A number of SMC samplers have been developed in the ABC framework (see \shortciteNP{sisson+ft07}, \shortciteNP{toni+wsis09}, \shortciteNP{beaumont+cmr09}, \shortciteNP{drovandi2011estimation}, \shortciteNP{delmoral+dj12}). 
Algorithm 8 presents a generalisation (to general kernels $K_h$) of the adaptive ABC SMC sampler of \shortciteN{delmoral+dj12}.

This algorithm provides an alternative method to rejection control  to avoid the collapse of the particle population, for an arbitrary choice of kernel $K_h$, by making particular sampler design choices.
Firstly, the probability of generating particles $\theta_m^{(i)}$ with identically zero weights $w^{(i)}_m=0$ is reduced by
increasing the number of summary statistics drawn to $T$, thereby targeting the joint distribution $\pi_{ABC}(\theta,s(1),\ldots,s(T)|s_{obs})$ as described in Section \ref{sec:pseudo-marginal}, although at the price of greater computation. Within the scope of an ABC SMC sampler that makes use of MCMC kernels within the move step (as with Algorithm 8), the alternative algorithms analysed by \citeN{lee+l14} (see Section \ref{sec:alternative}) could also be implemented (e.g. \shortciteNP{bernron+jgr17}).

In combination with the increased number of summary statistic replicates, Algorithm 8 directly controls the degree of particle degeneracy in moving from distribution $f_{m-1}(\theta)$ to $f_m(\theta)$. In particular, the next kernel scale parameter $h_m<h_{m-1}$ is chosen as the value which results in the effective sample size following the reweighting step, being reduced by a user specified proportion, $\alpha$. In this manner, the sample degeneracy will reduce in a controlled manner at each iteration, and the sequence of $h_m$ will adaptively reduce at exactly the rate needed to achieve this. When the effective sample size is reduced below some value $E$, resampling occurs and resets the effective sample size back to $N$, and the process repeats. As a result, resampling repeatedly occurs automatically after a fixed number of reweighting steps, as determined by $\alpha$.

This algorithm requires a stopping rule to terminate. If left to continue, $h_m$ would eventually  reduce very slowly, which is an indication that the sampler can no longer efficiently move the particles around the parameter space. \shortciteN{delmoral+dj12} argue that this identifies natural values of $h_M$ that should then be adopted. In particular, they terminate their algorithm when the MCMC move rate drops below 1.5\%, which then determines the final value of $h_M$. Alternative strategies to adaptively choose the kernel scale parameter sequence have been proposed by \citeN{drovandi2011estimation},  \citeN{drovandi2011likelihood}, \shortciteN{silk+fs13} and \shortciteN{daly+ghc17}.

SMC algorithms provide many easy opportunities for sampler adaptation, unlike MCMC samplers which are constrained by the need to maintain the target distribution of the chain. For example, within ABC SMC algorithms, 
\citeN{prangle17} adaptively learns the relative weightings of the summary statistics within the distance function $\|s-s_{obs}\|$ to improve efficiency, 
\citeN{bonassi+w15} construct adaptive move proposal kernels $g_m(\theta_{m}^{(i)},\theta)=\sum_{i=1}^N\nu^{(i)}_{m-1}F_m(\theta_{m-1}^{(i)},\theta)$ based on weighting components of $g_m$, via $\nu^{(i)}_{m-1}$, based on the proximity of $s^{(i)}_{m-1}$ to $s_{obs}$,
and \shortciteN{filippi+bcs13} develop a different method of adaptively constructing the sequence of intermediary distributions, $f_m(\theta)$,  based on Kullback-Leibler divergences between successive distributions.

Other ideas can be incorporated within ABC SMC algorithms in particular settings, or can use the ideas from ABC SMC algorithms to tackle problems related to posterior simulation.
For example, \shortciteN{prangle+ek17} use ideas from rare event modelling to improve sampler efficiency within ABC SMC algorithms.
When simulation from the model $p(s|\theta)$ is expensive, \citeN{everitt+r17} first use a cheap approximate simulator within an ABC SMC algorithm to rule out unlikely areas of the parameter space, so that expensive computation with the full simulator is avoided until absolutely necessary.
\shortciteN{jasra+smm12} implement an ABC approximation within an SMC algorithm to perform filtering for a hidden Markov model.
\shortciteN{dean+sjp14} and \shortciteN{yildirim+sdj15} use ABC SMC methods for optimisation purposes (with a different sequence of intermediary distributions), so as to derive maximum (intractable) likelihood estimators for hidden Markov models.

\begin{table}[tbh]
\caption{\bf Algorithm 8: ABC Sequential Monte Carlo Algorithm}
\noindent {\it Inputs:}
\begin{itemize}
\item A target posterior density $\pi(\theta|y_{obs})\propto p(y_{obs}|\theta)\pi(\theta)$, consisting of a prior distribution $\pi(\theta)$ and a procedure for generating data under the model $p(y_{obs}|\theta)$.
\item A kernel function $K_h(u)$, and an integer $N>0$.
\item An initial sampling density $g(\theta)$ and sequence of proposal densities $g_m(\theta,\theta')$, $m=1,\ldots,M$.
\item A value $\alpha\in[0,1]$ to control the effective sample size.
\item A low dimensional vector of summary statistics $s=S(y)$.
\end{itemize}

\noindent {\it Initialise:}\\
For $i=1,\ldots,N$: 
\begin{itemize}
\item Generate $\theta^{(i)}_0 \sim g(\theta)$ from initial sampling distribution $g$.
\item Generate $y^{(i)}_0(t) \sim p(y|\theta^{(i)}_0)$ and compute summary statistics $s^{(i)}_0(t)=S(y^{(i)}_0)$ for $t=1,\ldots,T$.
\item Compute weights 
$w^{(i)}_0= 
\pi(\theta^0_i)/g(\theta^{(i)}_0)$, and set $m=1$.
\end{itemize}

\noindent {\it Sampling:}
\begin{enumerate}
\item \label{alg8:step1} Reweight: Determine $h_m$ such that $ESS(w^{(1)}_m,\ldots,w^{(N)}_m)=\alpha ESS(w^{(1)}_{m-1},\ldots,w^{(N)}_{m-1})$
where  
\[	w^{(i)}_m=w^{(i)}_{m-1}\frac{\sum_{t=1}^TK_{h_m}(\|s^{(i)}_{m-1}(t)-s_{obs}\|)\pi(\theta^{(i)}_m)}
	{\sum_{t=1}^TK_{h_{m-1}}(\|s^{(i)}_{m-1}(t)-s_{obs}\|)\pi(\theta^{(i)}_{m-1})},
\]
and then compute new particle weights and set $\theta^{(i)}_m=\theta^{(i)}_{m-1}$ and $s^{(i)}_m(t)=s^{(i)}_{m-1}(t)$  for $i=1,\ldots,N$, and $t=1,\ldots,T$.

\item \label{alg8:step2} Resample: If $ESS(w^{(1)}_m,\ldots,w^{(N)}_m)<E$ then resample $N$ particles from the empirical distribution function $\{\theta^{(i)}_m,s^{(i)}_m(1),\ldots,s^{(i)}_m(T), W_m^{(i)}\}$ where $W_m^{(i)}=w_m^{(i)}/\sum_{j=1}^Nw_m^{(j)}$ and set $w_m^{(i)}=1/N$.

\item \label{alg8:step 3} Move: For $i=1,\ldots,N$: If $w_m^{(i)}>0$:
\begin{itemize}
\item Generate $\theta'\sim g_m(\theta^{(i)}_m,\theta)$, $y'(t)\sim p(y|\theta^{(i)}_m)$ 
and compute $s'(t)=S(y'(t))$ for $t=1,\ldots,T$.

\item Accept $\theta'$ with probability
\[
	\min\left\{1,\frac{\sum_{t=1}^TK_{h_m}(\|s'(t)-s_{obs}\|)\pi(\theta')g(\theta',\theta^{(i)}_m)}
	{\sum_{t=1}^TK_{h_{m}}(\|s^{(i)}_{m}(t)-s_{obs}\|)\pi(\theta^{(i)}_{m})g(\theta^{(i)}_m,\theta')}\right\}
\]
and set $\theta_m^{(i)}=\theta'$, $s^{(i)}_m(t)=s'(t)$ for $t=1,\ldots,T$.

\end{itemize}
\item Increment $m=m+1$. If stopping rule is not satisfied, go to \ref{alg8:step1}.
\end{enumerate}

\noindent {\it Output:}\\
A set of weighted parameter vectors $(\theta^{(1)}_M,w^{(1)}_M),\ldots,(\theta^{(N)}_M,w^{(N)}_M)$ drawn from $\pi_{ABC}(\theta|s_{obs})\propto \int K_{h_M}(\|s-s_{obs}\|)p(s|\theta)\pi(\theta)ds$.
\end{table}

\section{Discussion}
\label{sec:other}

ABC samplers have proved to be highly accessible and simple to implement, and it is this that has driven the popularity and spread of ABC methods more generally.
Multi-model versions of each of these algorithms are available (e.g. \shortciteNP{toni+wsis09}, \shortciteNP{chkerbtii+ccb15}) or can be easily constructed, with ABC posterior model probabilities and Bayes factors being determined by the relative values of $\frac{1}{N}\sum_{i=1}^NK_h(\|s^{(i)}-s_{obs}\|)$ under each model.
Although here the user needs to clearly understand the ideas of summary statistic informativeness for model choice (e.g. \shortciteNP{marin+prr14}) and the problems involved in computing Bayes factors as $h\rightarrow 0$ \shortcite{martin+frr17}.

Improvements to general ABC samplers include increasing algorithmic efficiency by using quasi Monte Carlo methods \cite{buchholz+c17}, and the use of multi-level rejection sampling \shortcite{warne+bs17} (see \shortciteNP{jasra+jnst17} for the SMC version) for variance reduction.
The lazy ABC method of \citeN{prangle16} states that it may be possible to terminate expensive simulations $s\sim p(s|\theta)$ early, if it is also possible to calculate the probability that the full simulation when run to completion would have been rejected.
Diagnostics to determine whether the kernel scale parameter $h_m$ is sufficiently low that $\pi_{ABC}(\theta|s_{obs})$ is indistinguishable from $\pi(\theta|s_{obs})$ were developed by \shortciteN{prangle+bps14}.
There are many related results on the rate of convergence of ABC algorithms as measured through the mean squared error of point estimates \shortcite{blum10,fearnhead+p12,calvet+c15,biau+cg15,barber+vm15}.
ABC samplers have also allowed previously unclear links to other algorithms to become better understood -- for example, \shortciteN{nott+mt12} have reinterpreted the Kalman filter as an ABC algorithm, and \shortciteN{drovandi18} (this volume) has comprehensively described the links between ABC and indirect inference.

A number of algorithms related to ABC methods have emerged, including 
Bayesian empirical likelihoods \shortcite{mengersen+pr13} and bootstrap likelihoods \shortcite{zhu+ml16}, the synthetic likelihood (\citeNP{wood10}, \shortciteNP{drovandi+gkr18}, this volume), the expectation-propagation ABC algorithm (\shortciteNP{barthelmec14,barthelme+cc17}, this volume), 
\shortciteN{albert+ks15}'s particle-based simulated annealing algorithm, 
and \citeN{forneron+n16}'s
 optimisation-based likelihood free importance sampling algorithm.
Perhaps the biggest offshoot of ABC samplers is the more general pseudo-marginal Monte Carlo method \cite{beaumont03,andrieu+r09}, which implements exact Monte Carlo simulation with an unbiased estimate of the target distribution, of which ABC is a particular case. See \shortciteN{andrieu+sv17} (this volume) for an ABC-centred exploration of these methods.

\section*{Acknowledgements}

SAS is supported by the Australian Research Council under the Discovery Project scheme (DP160102544), and the Australian Centre of Excellence in Mathematical and Statistical Frontiers (CE140100049).

\bibliographystyle{chicago}
\bibliography{biblio}
\thispagestyle{empty}

\end{document}